# Stable single photon source in near infrared

T. Gaebel, I. Popa, A. Gruber, M. Domhan, F. Jelezko and J. Wrachtrup

*3. Physikalisches Institut, Universität Stuttgart, Pfaffenwaldring 57, 70569 Stuttgart, Germany*

**Abstract**

Owing to their unsurpassed photostability, defects in solids may be ideal candidates for single photon sources. Here we report on generation of single photons by optical excitation of a yet unexplored defect in diamond, the nickel-nitrogen complex (NE8) centre. The most striking feature of the defect is its emission bandwidth of 1.2 nm at room temperature. The emission wavelength of the defect is around 800 nm, which is suitable for telecom fibres. In addition, in this spectral region little background light from the diamond bulk material is detected. Consequently, a high contrast in antibunching measurements is achieved.

**Introduction**

The technology for creation of single photons is attracting permanent attention because of possible applications in quantum cryptography and computation [1]. Several schemes to produce single photons on demand have been proposed ranging from faint laser pulses to optically excited impurities in solids [2] and quantum dots. Single quantum systems show important advantages in comparison to attenuated classical (poissonian) sources because of the complete suppression of multiple photon emission events. This property of single quantum state emission is related to the influence of the measurement process on the state of a single quantum system. When a photon emitted by, e.g., a single molecule, is detected, the system is projected into the ground electronic state. Hence, the emission of a second simultaneous photon is impossible. In the recent past, several of these systems have been proposed as single photon emitters. First attempts have been carried out on single molecules at low temperature [3] as well as under ambient condition [4,5]. In the latter case, a major drawback is the limited photostability of organic dye molecules at room temperature, whereas in the first case, liquid helium temperatures are prohibitive for most practical applications. Nevertheless, molecules are still interesting candidates for single photon sources mainly because they allow for flexibility in the choice of emission wavelength. Alternatively, quantum dots have been proposed as emitters. For this case, electrical and optical pumping of single photon emission had been shown [6,7,8]. A certain disadvantage is the limited emission wavelength and the requirement of low temperature operation. The only photostable single photon source at room temperature reported so far is the nitrogen-vacancy defect centre in diamond [9,10,11]. A serious disadvantage of the N-V defect is its spectrally broad emission band and also the emission wavelength, which is around 640 nm.

It is known that certain types of defects in diamond, in particular when they are not associated with a vacancy, show a much narrower emission range [12]. In the following we will describe such a defect. Recently several new defects emitting in the infrared spectral region which are related to Ni ions have been described in literature [13,14,15]. It should be noted that the absorption of the Ni-related centres is more than one order of magnitude weaker in natural than in synthetic diamonds. This is because untreated stones contain defects related to impurity ions present in the crystal lattice in very low concentration.

**Experimental**

For experiments, IIa type diamonds from Drukker were used. Single defects in diamond had been selected using a conventional confocal microscope. Excitation was carried out either by a pulsed laser (Coherent Mira; repetition frequency 76 MHz and excitation wavelength 695 nm) or continuously with an excitation wavelength of 710 nm. The fluorescence light from a single defect centre is collected and directed to the detection channel, which includes a spectrometer and a Hanbury-Brown and Twiss interferometer. Photon arrival events were analyzed using a single-photon counting card (SPCM 730, Becker&Hickl). An electronic delay was inserted in one of the detection channels in order to check the symmetry of the autocorrelation function. The fluorescence emission spectrum has been detected with a 0.3 m imaging spectrometer (Acton research) equipped with a back-illuminated CCD camera (Roper Scientific).

**Results and Discussion**

The fluorescence emission spectrum obtained from a single defect centre is shown in Fig. 2. The spectrum shows one pronounced zero-phonon line at 802 nm, together with a shallow phonon side wing extending up to 850 nm. The relative integral intensity of the zero phonon line to the entire spectrum (Debye-Waller factor) is 0.7, higher than the corresponding Debye-Waller factor for the N-V centre in diamond which is 0.04. Time resolved fluorescence decay measurements (data not shown) show a monoexponential decay curve with a decay constant of 11.5 ns.

Fluorescence intensity autocorrelation measurements have been performed to demonstrate the sub-poissonian statistics of the light emitted by the centre. Using a start-stop scheme the coincidence rate between the two detectors of a Hanbury-Brown and Twiss interferometer has been measured. Obtained data are equivalent to the second order intensity autocorrelation function for short time scales. The normalized autocorrelation function $g^{(2)}(\tau) = \frac{\langle I(t)I(t+\tau)\rangle}{\langle I(t)\rangle^2}$ was obtained from a photon coincidence rate histogram [16].

Data are presented in Fig. 3. The autocorrelation function shows a pronounced minimum at zero delay time. The remaining difference from zero at zero delay time results from the background, mostly related to the Raman scattering from the diamond lattice. In addition to antibunching, the autocorrelation function also shows photon bunching ($g^{(2)}(\tau) > 1$). Such a behavior was previously reported for single organic molecules and the N-V centre and is related to an additional metastable state in the photoexcitation cycle. Photon bunching is related to the lifetime of the metastable level and the intersystem crossing (ISC) rate. When recorded in time, the emission trace consists of bright (fluorescence) and dark (no fluorescence) intervals. The bright intervals correspond to the evolution of the system between the states 0 and 1 (see Fig. 1). If the system undergoes a transition to the metastable state, it will not fluoresce and a dark interval will occur with a length equal to the lifetime of the metastable state.

In order to determine the ISC rates, a theoretical model had been used to fit the data [17]. The evolution of the system can be described by optical Bloch equations for a three-level system (Fig. 1). In general, such a system can only be solved numerically. However, due to the high dephasing of the optical transition at room temperature (≈ 400 GHz), the optical Bloch equations can be reduced to rate equations and solved analytically.

$$\dot{\rho}_1 = -(\Gamma + k)\rho_1 + k\rho_0,$$
$$\dot{\rho}_0 = \left(\frac{1}{T_1} + k\right)\rho_1 - k\rho_0 + \gamma_{20}\rho_2, \qquad (1)$$
$$\dot{\rho}_2 = \Gamma_{12}\rho_1 - \gamma_{20}\rho_2,$$

where $\Gamma = \frac{1}{T_1} + \Gamma_{12}$, $\Gamma_{12}$ the ISC rate, $k$ the absorption rate and $\gamma_{20}$ the decay rate of the metastable state. The normalized autocorrelation function is $g^{(2)}(t) = \frac{p(t)}{p(\infty)}$, where $p(t)$ is the analytical solution for the two-photon correlator :

$$p(t) = \frac{k}{T_1}\left[\frac{\gamma_{20}}{\gamma_0^2 - R^2} + \left(1 - \frac{\gamma_{20}}{\gamma_0 - R}\right)\frac{e^{-(\gamma_0 - R)t}}{2R} - \left(1 - \frac{\gamma_{20}}{\gamma_0 + R}\right)\frac{e^{-(\gamma_0 + R)t}}{2R}\right], \qquad (2)$$

with $\gamma_0 = \frac{\Gamma + 2k + \gamma_{20}}{2}$ and $R = \sqrt{\left(\frac{\Gamma + 2k - \gamma_{20}}{2}\right)^2 - \Gamma_{12}k}$. The autocorrelation curves have been fitted with (2), using $\Gamma_{12}$ and $\gamma_{20}$ as fit parameters (Fig. 3). The fit values are shown in Table 1.

In order to determine the maximum emission rate and to derive the fluorescence quantum yield, the fluorescence saturation curve had been recorded (Fig.4). In the low laser power regime, the fluorescence intensity is linearly increasing with laser power. For high laser powers, the influence of the metastable level increases, and the fluorescence intensity saturates. The saturation data have been fitted using $R = R_\infty \frac{I/I_S}{1 + I/I_S}$, where $I_S$ is the laser power corresponding to saturation and $R_\infty$ is the fluorescence intensity corresponding to infinite laser power. With an obtained value for $R_\infty$ of 75000 counts/s, and with a detection efficiency of 0.5%, the count rate is $15 \times 10^6$ counts/s. The photon emission rate for this case is smaller than that for an ideal photon emitter, i.e., a two-level system. Using the ISC rates specified above, the maximum emission rate for a three-level system can be calculated [18]:

$$R_\infty = \frac{\left(\frac{1}{T_1} + \Gamma_{12}\right)\Phi_F}{2 + (\Gamma_{12}/\gamma_{20})}, \qquad (3)$$

where $\Phi_F$ is the fluorescence quantum yield. We estimate that our detection efficiency is around 0.5% which results in a $\Phi_F = 0.7$. For $\Gamma_{12}$ = 0, a two-level maximum rate is obtained. With the rates obtained (see Table 1), the ratio between the maximum emission rates for three- and two-level systems is 0.5, i.e. the measured value for the fluorescence intensity is only 50% of the maximum possible. This high value is due to the fact that the ISC rate is low compared to the radiative emission rate. In addition, the decay rate of the metastable state is high, hence, the photon counting rate is comparable to that of the N-V centre.

**Conclusions**

We report on a new candidate for a single photon source based on single NE8 defect in diamond. This defect shows an intense and spectrally narrow emission, suitable for generation of transform-limited single photon wavepackets, which is important for optical quantum computing schemes [19]. Further more, the emission is in the near infrared, making it important for fibre-based quantum cryptography schemes.

**Acknowledgement**

This work is supported by the DFG Schwerpunktprogramm "Quanten-Informationsverarbeitung".

**Figure captions**

Fig. 1: Left: NE8 structure consisting of one Ni atom surrounded by four N atoms. Right: the energy levels scheme of NE8; corresponding decay and ISC rates are shown. The ground and first excited states are labeled with 0 and 1, respectively, while the metastable state is labeled with 2.

Fig. 2: The fluorescence spectrum for the NE8 defect shows a narrow (1.2 nm) zero-phonon line at 802 nm. The weak phonon side wing results in the high Debye-Waller factor. The Raman scattering peak is at 780 nm.

Fig. 3: The fluorescence intensity autocorrelation function; the experimental data are shown in circle-shaped symbols, while the red line is the theoretical curve, calculated according to the procedure described in the text. The autocorrelation function shows a good contrast and presents a pronounced dip close to t=0.

Fig. 4: The fluorescence intensity saturation curve; symbols represent the experimental data and the red line a fit to the data, according to the formula in the text.

Table1: The fit parameter values from the autocorrelation function

| Symbols | Meaning | Value [MHz] |
|---|---|---|
| $k$ | Absorption rate | 440 |
| $\frac{1}{T_1}$ | Relaxation rate of the excited state | 87 |
| $\Gamma_{12}$ | ISC rate | 17 |
| $\gamma_{20}$ | Decay rate of the metastable state | 6.1 |

Fig. 1:

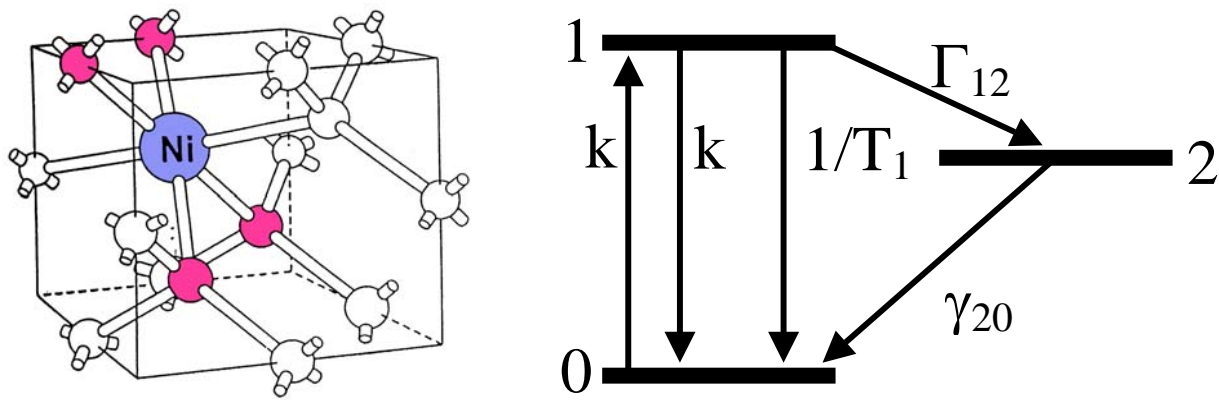

Fig. 2:

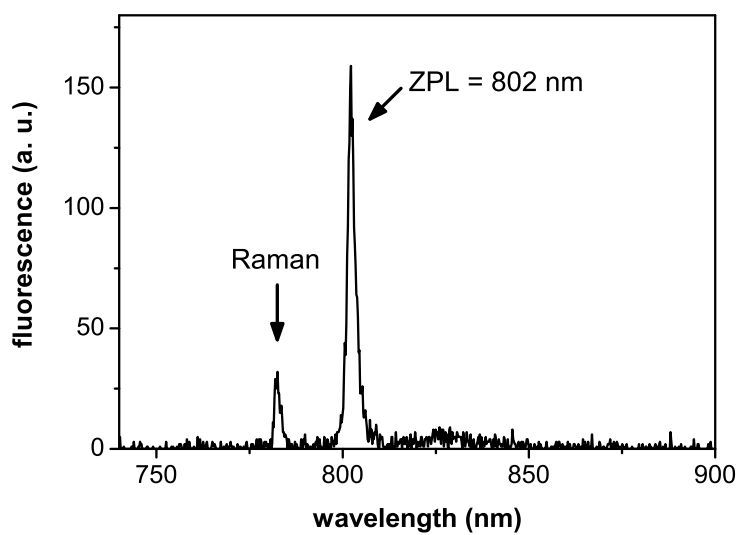

Fig. 3:

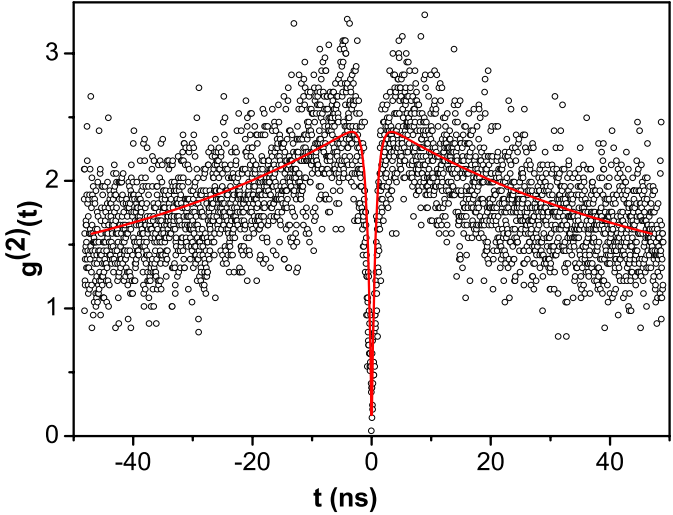

Fig. 4:

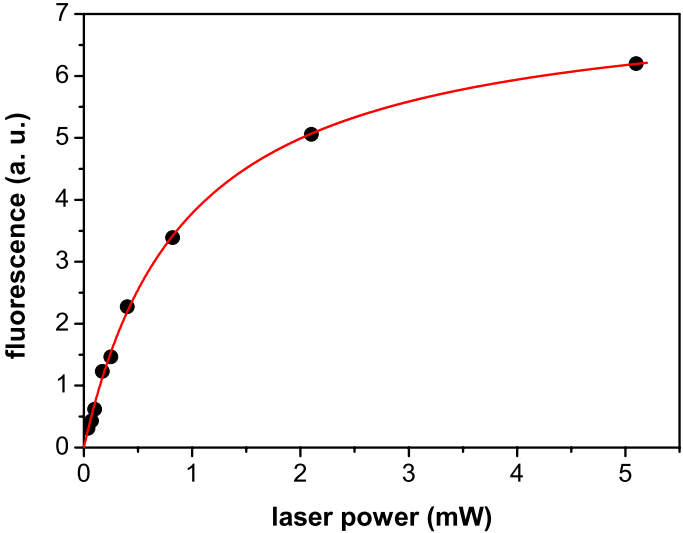